# SELECTIVE CONTROL OF RECONFIGURABLE CHIRAL PLASMONIC METAMOLECULES

# Anton Kuzyk[1,2,*,†], Maximilian J. Urban[1,3,†], Andrea Idili[4,†], Francesco Ricci[4,*], Na Liu[1,3,*]

[1] Max Planck Institute for Intelligent Systems, Heisenbergstrasse 3, D-70569 Stuttgart, Germany.
[2] Department of Neuroscience and Biomedical Engineering, Aalto University School of Science, P.O. Box 12200, FI-00076 Aalto, Finland.
[3] Kirchhoff Institute for Physics, University of Heidelberg, Im Neuenheimer Feld 227, D-69120 Heidelberg, Germany.
[4] Chemistry Department, University of Rome Tor Vergata, Via della Ricerca Scientifica, Rome 00133, Italy.

[*] Corresponding author. E-mail: anton.kuzyk@aalto.fi (A.K.); francesco.ricci@uniroma2.it (F.R.); laura.liu@is.mpg.de (N.L).
[†] These authors contributed equally to this work

## Abstract

Selective configuration control of plasmonic nanostructures has remained challenging using both top-down and bottom-up approaches in the field of active plasmonics. In this Article, we demonstrate the realization of DNA-assembled reconfigurable plasmonic metamolecules which can respond to a wide range of pH changes in a programmable manner. Such programmability allows for selective reconfiguration of different plasmonic metamolecule species coexisting in solution through simple pH tuning. Importantly, this approach enables discrimination of chiral plasmonic quasi-enantiomers as well as arbitrary tuning of chiroptical effects with unprecedented degrees of freedom. Our work outlines a new blueprint for implementation of advanced active plasmonic systems, in which individual structural species can be programmed to perform multiple tasks and functions.

## Introduction

Active control of plasmonic metamolecules is a burgeoning new direction in nanoplasmonics, which holds great promises for realization of novel active devices, such as optical switches, transducers, modulators, filters, phase shifters at different wavelengths(*1*). In general, there are two major schemes to implement active plasmonic systems. One is based on integration of active media, i.e., liquid crystals, phase change materials, III–V semiconductors, whose material properties can be altered upon external stimuli(*2–7*). The other is based on geometrical reconfiguration(*8, 9*), i.e. structural tuning of plasmonic metamolecules. This latter scheme requires controllable actuation at the nanoscale, which often encounters substantial constraints and technological challenges.

At optical wavelengths, one unique technique for structural tuning of plasmonic nanostructures is DNA self-assembly, taking advantage of the inherent molecular recognition and programmability of DNA chemistry(*10–13*). This approach allows for large-scale production of plasmonic nanostructures in a highly parallel manner(*14–18*). Importantly, it enables reconfigurable plasmonic systems, which are beyond the state of the art of top-down nanofabrication techniques(*8, 19, 20*). There are various ways to control the active behavior of plasmonic nanostructures through DNA self-assembly. Probably, the most versatile and thus widespread approach is based on the so-called "toehold-mediated strand displacement reaction"(*21*), which utilizes DNA strands as fuel to regulate spatial configuration(*22–24*). Photoresponsive molecules such as azobenzene can also be employed through incorporation with DNA to activate response upon light stimuli(*19, 25, 26*). More intriguing approaches could include reversible reconfiguration based on shape-complementarity(*27, 28*) or structural adaptions of aptamers to the presence of target molecules(*29*).

So far, only active control of plasmonic metamolecule ensembles that contain single structural species has been investigated. Selective control and active tuning of individual structural species coexisting within one ensemble has not been demonstrated. The power of programmable DNA chemistry for nanoplasmonics therefore has not yet been fully explored. Such selective activation of individual structural species could be prototypical building blocks for advanced plasmonic sensors, which can carry out multiple tasks and functions following a pre-designed "sense-activate" algorithm, as each structural species could be programmed to respond a specific set of inputs. In this Article, we experimentally demonstrate selective control of reconfigurable chiral plasmonic metamolecules assembled with DNA origami(*30, 31*). We utilize pH-sensitive DNA "locks" as active sites to trigger structural regulation of the chiral plasmonic metamolecules over a wide pH range. Such locks exploit triplex DNA secondary structures that display pH-dependent behavior due to the presence of specific protonation sites(*32, 33*). Crucial advantages of pH control over oligonucleotides lie in simpler implementation, faster modulation, as well as no generation of waste products, preventing system deteriorations. We further demonstrate discriminative reconfiguration of plasmonic quasi-enantiomers(*34*), achieving enantioselective tuning of chiroptical effects with unprecedented degrees of freedom.

**Results**

The working principle of a pH-sensitive DNA triplex formation is illustrated in Fig. 1A. A DNA triplex can be formed through pH-sensitive sequence specific parallel Hoogsteen interactions (dots) between a single-stranded DNA (ssDNA) and a duplex DNA, which is formed through pH-insensitive Watson-Crick interactions (dashed). While CGC triplets require protonation of the N3 of cytosine in the ssDNA (average p$K_a$ of protonated cytosines in a triplex structure is ~ 6.5) and thus are only stable at acidic pHs, TAT triplets are only destabilized at pHs above 10 due to the deprotonation of thymine (p$K_a$ ~ 10)(*32*, *35–38*). By varying the relative content of TAT/CGC triplets, it is thus possible to design DNA locks that can be opened or closed over specific pH windows(*32*).

By implementing such DNA locks on reconfigurable DNA origami, pH-triggered structural tuning of chiral plasmonic metamolecules can be achieved (see Fig. 1B). Each metamolecule comprises two gold nanorods (AuNRs) assembled on a cross-like DNA origami template (see figs. S1-S3 and tables S1-S3 in Supplementary Materials for details of the structural design and the assembly process). The DNA lock is composed of a 20-basepair duplex and a 20-base ssDNA positioned on the two respective bundles of the template such that the pH-regulated triplex formation triggers a conformational change of the metamolecule from a "relaxed" state to a "locked" state (Fig. 1B). Of note, the chiral plasmonic nanostructures can be switched to the left-handed (LH) or the right-handed (RH) state depending on the initial positions of the duplex and the ssDNA on the origami (see Fig. 1B) over different pH windows. The angle between the AuNRs in both the LH and RH configurations is designed to be ~50°.

Fig. 2A presents a transmission electron microscopy (TEM) image of the DNA origami/AuNR hybrid nanostructures (see also fig. S4). A high assembly yield has been achieved. Circular dichroism (CD), that is, differential absorption of LH and RH circularly polarized light by chiral structures, is used as a measure to optically characterize our samples. Representative CD spectra of the metamolecules in the LH and RH states are shown by the red and blue curves in Fig. 2B, respectively. The pH-sensitive DNA lock used in these nanostructures contains equal amounts of TAT and CGC triplets (50% TAT). The CD spectra exhibit characteristic bisignate profiles (*39–41*).

First, we demonstrate that the reconfiguration of our plasmonic metamolecules can be rationally controlled over a wide pH range by simply modulating the relative content of TAT/CGC triplets in the DNA lock (see Fig. 3A). To demonstrate the pH-dependent behavior, the CD responses of the plasmonic metamolecules in buffers of different pHs are measured at 740 nm (Fig. 2b, dashed line) using a CD spectrometer (Jasco-815). As shown in Fig. 3B, when a DNA lock that contains 50% TAT is employed, due to the stabilization of the DNA triplex through the Hoogsteen interactions, the plasmonic nanostructures are mostly locked in the LH or RH state at pH below 6.5. The triplex-to-duplex transition in the DNA lock occurs upon a pH increase. For both the LH and RH structures, gradual amplitude decreases in CD response are observed, until it approximately reaches zero at pH ~7, indicating that the plasmonic metamolecules are transformed into the relaxed state. The pH of semiprotonation p$K_a^{obs}$ (the average p$K_a$ due to several interacting protonation cites) for the metamolecules functionalized with 50% TAT locks is ~ 6.6. When the TAT content is as high as 80% in the DNA lock, the plasmonic nanostructures are opened at a much higher pH (p$K_a^{obs}$ =8.7). The cases of 60% and 70% TAT contents show switching at intermediate pH values with p$K_a^{obs}$ of 7.2 and 7.8, respectively (see Fig. 3C).

Our plasmonic metamolecules exhibit fast response to pH changes. The time-dependent CD response of an exemplary sample (70% TAT) (Fig. 3D) shows that the reconfiguration upon a dynamic pH change between 9.5 and 6.0 takes place in a few minutes. Reconfiguration is significantly faster than the previous DNA origami-based reconfigurable plasmonic systems(*8*, *19*), which had switching kinetics on the order of several tens of minutes. For those plasmonic systems driven by DNA fuel strands (*8*, *20*) and azobenzene photoisomerization (*19*), the reconfiguration rates were largely limited by the kinetics of strand displacement reactions and azobenzene-modified DNA hybridization/dehybridization processes, respectively. In our pH-triggered plasmonic system the high effective concentration of the two triplex-forming domains enables a faster reconfiguration process, which is mainly limited by the rate of protonation/deprotonation of the cytosines/thymines in the ssDNA.

The possibility to engineer plasmonic metamolecules with programmable pH-dependent configurations opens a pathway towards enantioselective control of chiral plasmonic nanostructures. To demonstrate this, we mix LH and RH metamolecules in equimolar amounts (Fig. 4A). These LH and RH nanostructures are plasmonic quasi-enantiomers(*34*), that is, they are enantiomers as plasmonic objects, but functionalized with DNA locks containing different TAT contents. In the first experiment, only one LH group and one RH group are utilized. The pH-dependence result (black curve) in Fig. 4B is obtained from a plasmonic quasi-racemic solution, in which the TAT contents in the RH and LH nanostructures are 50% (RH 50%; p$K_a^{obs}$ =6.6) and 60% (LH 60%; p$K_a^{obs}$ =7.2), respectively. At pH 5.5, the plasmonic mixture, which is nominally quasi-racemic, exhibits slightly nonzero CD, likely due to structural imperfections. In the range of pH [5.5, 6.5], where both the LH and RH structures are mainly stable in their locked states, the CD response of the mixture does not experience significant variations upon pH change. When the pH further increases, RH 50% (p$K_a^{obs}$ =6.6) starts to respond and the RH structures in the solution are gradually opened, going to the relaxed state. This is reflected by a substantial CD response decrease of RH 50% (see Fig. 3B). In contrast, LH 60% with p$K_a^{obs}$ =7.2 remains unresponsive in the solution. As a result, an abrupt CD peak is observed at pH 7.0 (Fig. 4B, black curve). This elucidates that discriminative reconfiguration of the plasmonic quasi-enantiomers has been successfully carried out. When the pH continues to increase, LH 60% becomes also responsive and the LH structures in the solution are opened gradually. The overall CD response of the mixture therefore decreases until pH 8. Subsequent pH increases do not introduce drastic CD changes in that both the LH and RH structures are in the relaxed state. Similarly, the quasi-racemic solution containing LH 50% and RH 60% exhibits nearly mirrored pH-dependence (Fig. 4B, grey curve).

To demonstrate the unprecedented degrees of freedom of our enantioselective control, four groups of metamolecules (two RH and two LH) are utilized in the second experiment. The pH-dependent CD result (Fig. 4C, black curve) is obtained from a racemic solution, containing RH 50% ($pK_a^{obs}$ =6.6), LH 60% ($pK_a^{obs}$ =7.2), RH 70% ($pK_a^{obs}$ =7.8), and LH 80% ($pK_a^{obs}$ =8.6) in equimolar amounts. The experimental curve in the range of pH [5.5, 7.5] resembles the black curve in Fig. 4B. In this range, only RH 50% and LH 60% are pH-triggered, whereas RH 70% and LH 80% are mainly unresponsive to pH change (see Fig. 3B) and remain stable in their respective locked states. When the pH subsequently increases, RH 70% and LH 80% start to undergo structural reconfiguration, transitioning to the relaxed state. On the contrary, RH 50% and LH 60% are already in the relaxed state, being pH unresponsive in the range of pH [7.5, 9.5]. A new CD peak is observed around pH 8.5. This is due to the fact that RH 70% responds to lower pHs than LH 80%. As a result, the RH structures in RH 70% are gradually opened, whereas LH 80% can remain in the locked state until pH 8.5 (see Fig. 3B). When the pH further increases, the LH structures in LH 80% are also gradually opened. Eventually, all four groups are in the relaxed state, resulting in a negligible CD response. The pH-dependent CD result of another sample, which contains RH 60%, LH 50%, RH 70%, and LH 80% in equimolar amounts is presented by the grey curve in Fig. 4C. In the range of pH [5.5, 7.5], the pH-dependent CD curve is nearly mirrored to the black curve, similar to the cases in Fig. 4B. In the range of pH [7.5, 9.5], the grey curve is nearly identical to the black curve. These results elucidate excellent control selectivity of different plasmonic nanostructure species using simple pH tuning over a wide range. In turn, our strategy outlines an innovative approach to arbitrarily modulate chiroptical response at wish.

**Discussion**

In conclusion, we have demonstrated selective control of reconfigurable chiral plasmonic metamolecules assembled by DNA origami. The pH-dependent reconfiguration of the plasmonic metamolecules can be tuned in a programmable way through adjusting the relative content of TAT/CGC triplets in the pH-sensitive DNA locks. Excellent enantioselectivity and low cross-activation of the plasmonic enantiomers in quasi-racemic mixtures have been demonstrated upon pH tuning over a wide pH range. Our pH-triggered plasmonic systems could be particularly suitable for optical monitoring pH-changes in biochemical media, plasmonic sensing of dynamic processes, and smart nanomechanical devices.

## Materials and Methods

### Materials

DNA scaffold strands (p7650) were purchased from Tilibit Nanosystems. Unmodified staple strands (purification: desalting) were purchased from Eurofins MWG. Capture strands for the AuNRs (purification: desalting) and DNA strands of pH-responsive locks were purchased from Sigma-Aldrich. Thiol-modified strands (purification: HPLC) were purchased from biomers.net. Agarose for electrophoresis and SYBR Green nucleic acid stain were purchased from ThermoFischer. Uranyl formate for negative TEM staining was purchased from Polysciences, Inc. AuNRs (10 nm × 38 nm) were purchased from Sigma-Aldrich (catalogue no. 716812). Other chemicals were purchased either from Carl-Roth or from Sigma-Aldrich.

### Design and assembly of DNA origami templates

DNA origami structures were designed using caDNAno v2.0. The strand routing diagram of the origami structures can be found in fig. S1. The sequences of the staple strands and modifications used for pH-sensitive DNA locks are provided in tables S1 and S2. The origami structures were prepared by thermal annealing in a thermal cycling device (Eppendorf Mastercycler pro). For thermal annealing temperatures and times see table S3. All reaction mixtures contained: 10 nM of the p7650 scaffold, 100 nM of each staple including those modified with sequences for pH locks. Excess staple strands were removed by gel electrophoresis (1.5% agarose gels containing Sybr Green strain, 0.5×TBE and 11 mM $MgCl_2$). Target bands were cut out and the DNA origami template structures were extracted with Freeze `N Squeeze spin columns (BioRad).

### Assembly of metamolecules

The AuNR assembly procedure was adopted from previous studies(*8*, *19*). In short, functionalization of the AuNRs with thiolated DNA (SH-5' T16, biomers.net) was carried out following the low pH route(*42*, *43*). Excess thiolated DNA was removed by centrifugation. The purified AuNRs were added to the purified DNA origami template structures in an excess of 10 AuNRs per DNA origami structure. The mixture was annealed from 40°C to 20°C over 15 hours. After thermal annealing, second agarose gel purification step (0.5% agarose gel in 0.5× TBE buffer with 11 mM $MgCl_2$) was used to remove the excess AuNRs. DNA origami AuNR structures were extracted with Freeze `N' Squeeze spin columns (BioRad) and further centrifuged at 6000 rcf for 25 min. The supernatant was carefully removed and the metamolecules were dispersed in 0.5× TBE, 11 mM $MgCl_2$ 0.02% SDS (Sodium dodecyl sulfate). At this stage different metamolecule samples were diluted to the same concentration, i.e., the same absorption (OD~5, 10 mm light path) at the longitudinal plasmon resonance of the AuNRs (780 nm).

### TEM characterization

The DNA origami structures (with or without AuNRs) were imaged using a Philips CM 200 TEM operating at 200 kV. For imaging, the DNA origami structures (with or without AuNRs) were deposited on freshly glow-discharged carbon/formvar TEM grids (Science Services). The TEM grids were treated with a uranyl formate solution (0.75%) for negative staining of the DNA structures.

### pH regulation of plasmonic metamolecules

Metamolecules functionalized with different pH switches were first diluted at a concentration of ~2.5 nM in 0.5× TBE buffer with 11 mM $MgCl_2$ and 0.02% SDS (pH ~8.3). These stock solutions were further diluted for CD measurements. The dilution buffers contained 0.5× TBE, 11 mM $MgCl_2$, 0.02%

SDS and varying amounts of acetic acid or sodium hydroxide. The pH of the final metamolecule solutions varied between 5.5 and 9.5 with a step of 0.5 (measurement points in Fig. 3B and Figs. 4B, 4C). The dilution factors for each type of the metamolecules were 10 (Fig. 3B), 20 (Fig. 4B) and 40 (Fig. 4C). For characterization of the pH change kinetics (Fig. 3D), LH 70% metamolecules were initially dispersed in buffers with pH 6.5 and 9.5. Then sodium hydroxide or acetic acid was added to the solution to adjust the pH to 9.5 or 6.5, respectively.

**Optical characterizations**

The CD and ultraviolet–visible measurements were performed with a J-815 Circular Dichroism Spectrometer (Jasco) using Quartz SUPRASIL Ultra-Micro cuvettes (105.204-QS, Hellma Analytics) with a path length of 10 mm. Ultraviolet–visible absorption measurements were also performed with a BioSpectrometer (Eppendorf).

The CD response in dependence on pH for a single metamolecule group with different TAT contents (Fig. 3B) was analyzed by fitting the measurement results with the Hill function as follows:

$$CD = CD_{open} + \frac{(CD_{locked} - CD_{open}) \cdot [H^+]^n}{[H^+]^n + (K_a^{obs})^n}$$

Where $CD_{open}$ and $CD_{locked}$ represent the CD intensities of the metamolecules in the relaxed and locked states, respectively. $[H^+]$ represents the total concentration of the hydrogen ions ($[H^+]=10^{-pH}$), $K_a^{obs}$ is the observed acid constant of the metamolecules, and $n$ is the Hill coefficient. $pK_a^{obs}$ (Fig. 3C) was calculated using $pK_a^{obs} = -\log(K_a^{obs})$.

The kinetics of the metamolecules when undergoing the locked-to-relaxed transition (Fig. 3D left) was fitted well using a single exponential decay function, suggesting first-order reaction kinetics

$$CD(t) = CD_{open} + CD_{locked} \cdot e^{-(t-t_0) \cdot k}$$

Where $CD_{open}$ and $CD_{closed}$ represent the CD intensities of the metamolecules in the relaxed and locked states, respectively, and $t_o=120$ s.

In contrast, the kinetics of the metamolecules when undergoing the relaxed-to-locked transition (Fig. 3D right) was fitted well using a double exponential decay function.

$$CD(t) = CD_{locked} + CD_1 \cdot e^{-(t-t_0) \cdot k_1} + CD_2 \cdot e^{-(t-t_0) \cdot k_2}$$

Where $CD_1+CD_2=CD_{open}$ and $t_o=120$ s.

The CD response in dependence on pH for mixed metamolecule groups (Fig. 4) was fitted with the sum of two (Fig. 4B) or four (Fig. 4C) Hill functions. In this case, $n$ and $K_a^{obs}$ parameters for individual Hill functions were fixed with values obtained from the single metamolecule group results.

**Supplementary Materials**

Supplementary material for this article available at http://advances.sciencemag.org
Fig. S1. Scaffold/staple layout of the DNA origami template.
Table 1. Staple sequences of the DNA origami template.
Fig. S2. Schematics of the DNA origami template.
Table S2. DNA sequences of the triplex pH switches.
Table S3. Thermal annealing temperatures and times.
Fig. S3. TEM images of the DNA origami templates after thermal annealing and agarose gel purification.
Fig. S4. Additional TEM images of the DNA origami-based metamolecules.

**Acknowledgments:** We thank M. Kelsch for assistance with TEM. TEM images were collected at the Stuttgart Center for Electron Microscopy (StEM). **Funding:** N.L. was supported by the Sofja Kovalevskaja Award from the Alexander von Humboldt Foundation. A.K. was supported by a postdoctoral fellowship from the Alexander von Humboldt Foundation. N.L. were supported by a Marie Curie CIG Fellowship. We also thank for the financial support from the European Research Council (ERC) Starting Grant 'Dynamic Nano'. FR was supported by Associazione Italiana per la Ricerca sul

**Figures and Tables**

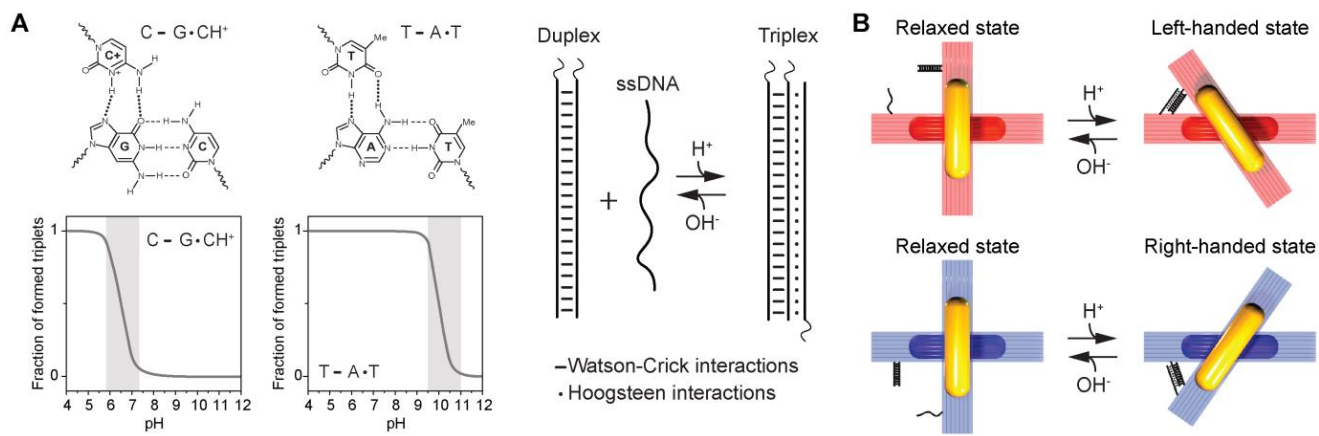

**Fig. 1. Working principle of the pH-sensitive plasmonic metamolecules.** (**A**) Left top: CGC and TAT triplets can be formed through the combination of Watson-Crick and Hoogsteen interactions. Left bottom: the formation of the CGC triplets requires protonation of cytosines and they are only stable at acidic pHs. In contrast, the TAT triplets are stable at pHs below 10 and unfold due to deprotonation of thymines. Right: pH-triggered DNA lock. (**B**) pH regulation of the DNA origami-based chiral metamolecules. The metamolecules can be switched between the relaxed and left-handed (LH)/right-handed (RH) state by opening/closing the pH-triggered DNA locks.

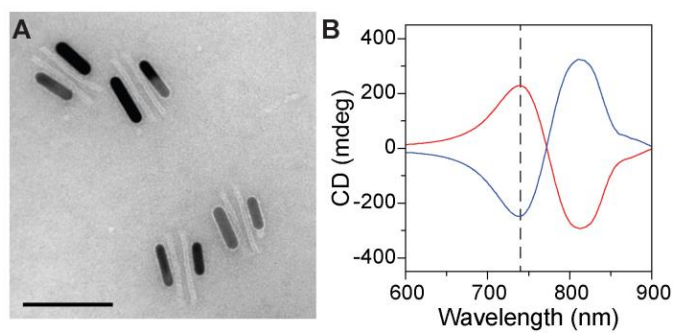

**Fig. 2.** DNA **assembly of the plasmonic metamolecules and their chirooptical response.** (**A**) TEM image of the plasmonic nanostructures. The structures tend to lie flat on the TEM grid. Scale bar, 100 nm. (**B**) Measured CD spectra of the metamolecules in the LH (red line) and the RH (blue line) state from the LH 50% and RH 50% samples at pH 5.5, respectively.

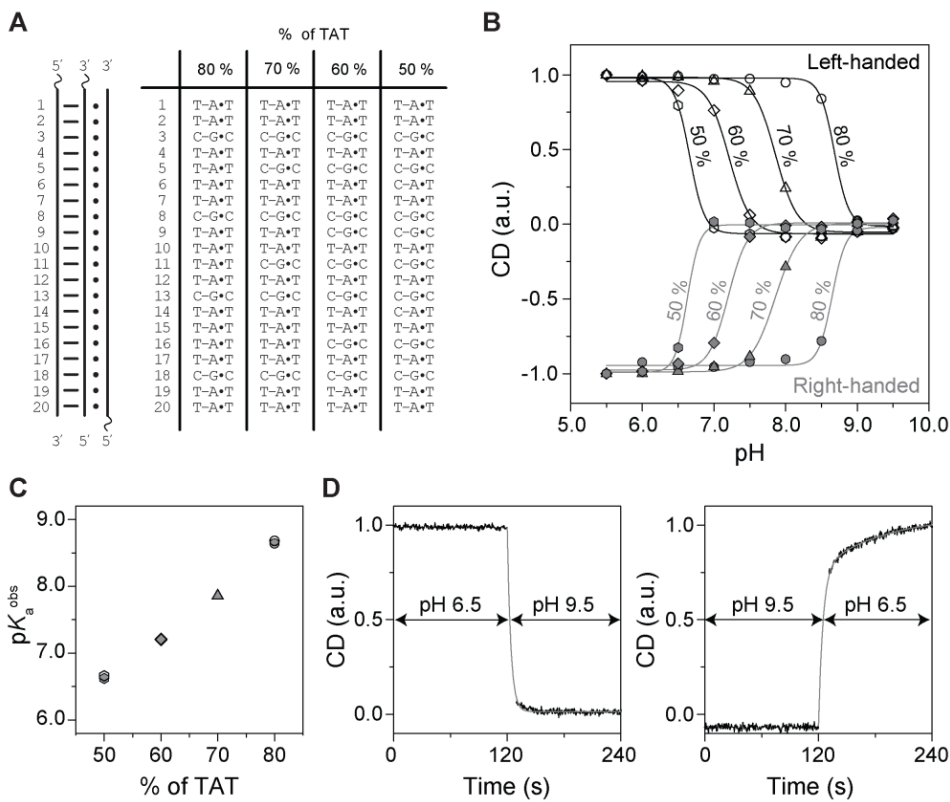

**Fig. 3. pH-dependent response of the plasmonic chiral metamolecules.** (**A**) Sequences of the DNA locks with different TAT/CGC contents. (**B**) Relative CD dependence on pH for the LH and RH metamolecules with DNA locks with different TAT contents. Solid lines are fitting results using Hill function (see Materials and Methods). (**C**) pH of semiprotonation (p$K_a^{obs}$) dependence on the TAT content for the LH (open black symbols) and RH (solid gray symbols) metamolecules. (**D**) Kinetic characterization of the metamolecules, switching between the LH state and the relaxed state upon pH changes.

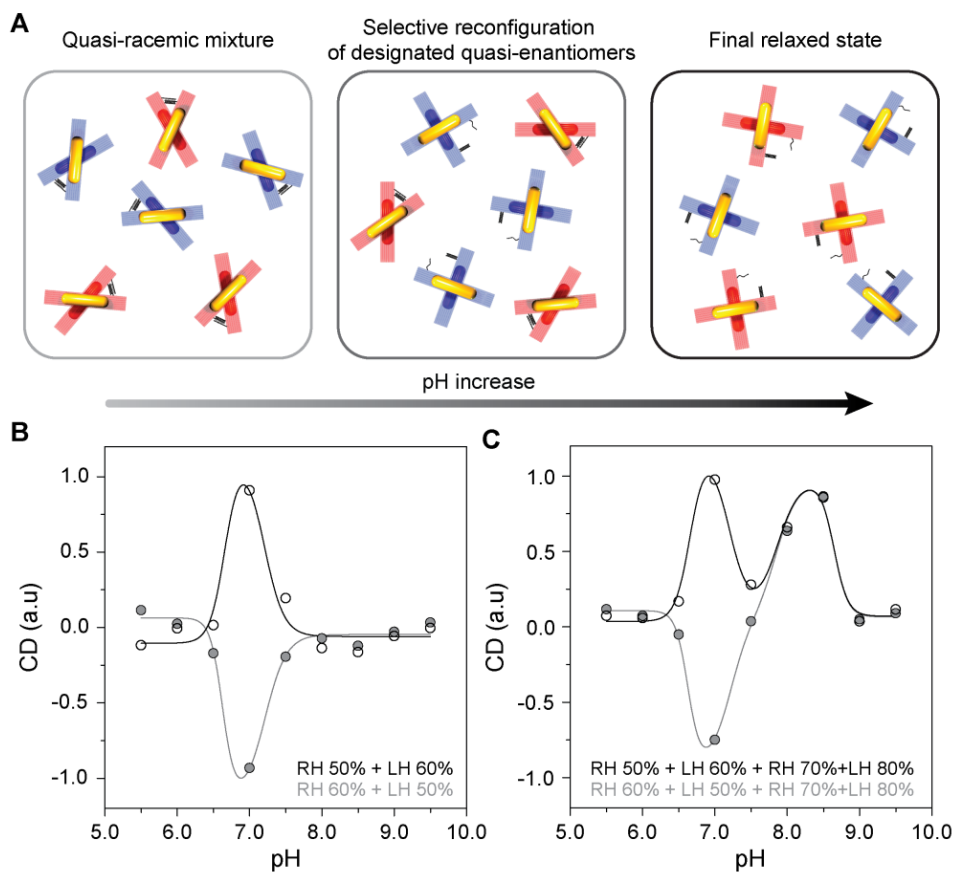

**Fig. 4. Enantioselective control of the chiral plasmonic metamolecules.** (**A**) Selective reconfiguration process of the designated quasi-enantiomers. Left: At low pHs, the LH (red) and RH (blue) metamolecules coexist in equimolar amounts. LH and RH metamolecules are quasi-enantiomers, i.e., they are enantiomers as plasmonics object but functionalized with DNA locks with different TAT content. The mixture is quasi-racemic. Middle: Upon a pH increase, one type of the quasi-enantiomers (blue) undergoes a locked to relaxed transition. Right: Further pH increases result in both quasi-enantiomers being in the relaxed state. (**B**) Relative CD dependence on pH for a mixture of two quasi-enantiomers, RH 50% and LH 60% (black open symbols); RH 60% and LH 50% (gray solid symbols). (**C**) Relative CD dependence on pH for a quasi-racemic mixture composed of four different metamolecules in equimolar amount, RH 50%, LH 60%, RH 70%, and LH 80% (black open symbols); RH 60%, LH 50%, RH 70% and LH 80 (gray solid symbols). Solid lines are fitting results from the sum of four Hill functions (see Materials and Methods).